\documentclass[journal]{IEEEtran}

\usepackage{xcolor,soul,framed} 

\colorlet{shadecolor}{yellow}
\usepackage[pdftex]{graphicx}
\graphicspath{{../pdf/}{../jpeg/}}
\DeclareGraphicsExtensions{.pdf,.jpeg,.png}

\usepackage[cmex10]{amsmath}
\usepackage{array}
\usepackage{mdwmath}
\usepackage{eqparbox}
\usepackage{url}
\usepackage{algorithm2e}
\usepackage{multirow}
\usepackage{listings}
\usepackage{graphicx}
\newcolumntype{P}[1]{>{\centering\arraybackslash}p{#1}}
\usepackage{amsfonts}
\usepackage{amsmath}
\usepackage{subfig}
\usepackage{ragged2e}
\usepackage{caption}
\usepackage{rotating}
\usepackage[style=base]{caption}
\begin{document}
\bstctlcite{IEEEexample:BSTcontrol}

    \title{Radiomic Deformation and Textural Heterogeneity (R-DepTH) Descriptor to characterize Tumor Field Effect: Application to Survival Prediction in Glioblastoma}
  \author{Marwa Ismail,
      Prateek Prasanna,
      Kaustav Bera,
      Volodymyr Statsevych,
      Virginia Hill,
      Gagandeep Singh,
      Sasan Partovi,
      Niha Beig,
      Sean McGarry,
      Peter Laviolette,
      Manmeet Ahluwalia,
      Anant Madabhushi,~\IEEEmembership{Fellow,~IEEE,}
      and~Pallavi Tiwari

\thanks{This work was funded by the NIH awards  
1U24CA199374-01, R01CA202752-01A1, R01CA208236-01A1, R01CA216579-01A1, R01CA220581-01A1, 1U01CA239055-01, 1U01CA248226-01, 1U54CA254566-01, 1P20CA233216-01;
National Institute for Biomedical Imaging and Bioengineering 1R43EB028736-01;
National Center for Research Resources under award number 1C06RR12463-01;
National Heart, Lung, and Blood Institute 1R01HL15127701A1;
VA Merit Review Award IBX004121A from the United States Department of Veterans Affairs; Biomedical Laboratory Research and Development Service;
the DOD Prostate Cancer Idea Development Award (W81XWH-15-1-0558);
the DOD Lung Cancer Investigator-Initiated Translational Research Award (W81XWH-18-1-0440);
the DOD Peer Reviewed Cancer Research Program (W81XWH-16-1-0329);
National Institute of Diabetes and Digestive and Kidney Diseases (1K25 DK115904-01A1);
the Ohio Third Frontier Technology Validation Fund;
the Wallace H. Coulter Foundation Program in the Department of Biomedical Engineering and the Clinical and Translational Science Award Program (CTSA) at Case Western Reserve University;
Department of Defense Peer Reviewed Cancer Research Program (PRCRP) Career Development Award;
Dana Foundation David Mahoney Neuroimaging Program;
The V Translational Cancer Research Foundation; America Brain Tumor Association DG1600004; NIH/NCI R01CA218144.}
  \thanks{M. Ismail, K. Bera, N. Beig, A. Madabhushi, and P. Tiwari are with the Biomedical Engineering department, Case Western Reserve University, OH (e-mail: mxi125@case.edu, kxb413@case.edu, ngb18@case.edu, axm788@case.edu, pxt130@case.edu).}
  \thanks{P. Prasanna is with the department of Biomedical Informatics, Stony Brook University, NY (e-mail: prateek.prasanna@stonybrook.edu).}%
  \thanks{V. Statsevych is with the imaging institute, Cleveland Clinic, OH (e-mail: statsev@ccf.org).}
  \thanks{V. Hill is with the department of Radiology, Northwestern University, IL (e-mail: virginiahill2@mac.com).}
  \thanks{G. Singh is with Radiology Department, Newark Beth Israel Medical Center, NJ (e-mail: gagan32092@gmail.com).}
  \thanks{S. Partovi is with Interventional Radiology Department, Cleveland Clinic, OH (e-mail: partovs@ccf.org).}
  \thanks{S. McGarry and P. Laviolette are with the Radiology Department, Medical College of Wisconsin, WI (e-mail: smcgarry@mcw.edu, plaviole@mcw.edu).}
  \thanks{M. Ahluwalia is with Brain Tumor and Neurooncology Center, Cleveland Clinic, OH (e-mail: ahluwam@ccf.org).}}


\maketitle

\begin{abstract}
The concept of tumor field effect implies that cancer is a systemic disease with its impact way beyond the visible tumor confines.  For instance, in Glioblastoma (GBM), an aggressive brain tumor, the increase in intracranial pressure due to tumor burden often leads to brain herniation and poor outcomes.  Our work is based on the rationale that highly aggressive tumors tend to grow uncontrollably, leading to pronounced biomechanical tissue deformations in the normal parenchyma, which when combined with local morphological differences in the tumor confines on MRI scans, will comprehensively capture tumor field effect. Specifically, we present an integrated MRI-based descriptor, radiomic-Deformation and Textural Heterogeneity (r-DepTH). This descriptor comprises measurements of the subtle perturbations in tissue deformations throughout the surrounding normal parenchyma due to mass effect. This involves non-rigidly aligning the patients’ MRI scans to a healthy atlas via diffeomorphic registration. The resulting inverse mapping is used to obtain the deformation field magnitudes in the normal parenchyma. These measurements are then combined with a 3D texture descriptor, Co-occurrence of Local Anisotropic Gradient Orientations (COLLAGE), which captures the morphological heterogeneity within the tumor confines, on MRI scans. R-DepTH, on N = 207 GBM cases (training set ($S_{t}$) = 128, testing set ($S_{v}$) = 79), demonstrated improved prognosis of overall survival by categorizing patients into low- (prolonged survival) and high-risk (poor survival) groups (on $S_{t}$, $p$-value = $3.5 \times 10^{-7}$, and on $S_{v}$, $p$-value = $0.0024$). R-DepTH descriptor may serve as a comprehensive MRI-based prognostic marker of disease aggressiveness and survival in solid tumors.  
\end{abstract}


%
\IEEEpeerreviewmaketitle


\section{Introduction}

\IEEEPARstart{I}{t} is well recognized that cancer is not a bounded, self-organized system \cite{chai2009field}, but a systemic disease. Most malignant tumors have heterogeneous growth patterns, leading to disorderly proliferation well beyond the visible surgical margins. In fact, in solid tumors, depending on the malignant phenotype, the impact of the tumor is observed not just within the visible tumor, but also in the immediate peri-tumoral, as well as in seemingly normal-appearing adjacent regions \cite{chai2009field,chan2018cancer}, a phenomenon known as `tumor field effect' \cite{lochhead2015etiologic}. For instance, Glioblastoma (GBM), one of the most aggressive brain tumors, is known to extend several millimeters distal to the tumor margins, which ultimately leads to recurrence in GBM patients \cite{salazar1976spread}. Similarly, the herniation or gross distortion of the brainstem, remote to the tumor location, is the proximal cause of deaths in 61\% of GBM  studies \cite{silbergeld1991cause}.  The infiltrating brain tumor mass pushes and displaces the surrounding tissue structures (known as mass effect), leading to a mid-line shift and an increase in the intracranial pressure \cite{tonn2006neuro,pan2003spinal}, which ultimately results in alterations of consciousness and other chronic conditions such as seizures and headaches in GBM patients. 
In a recent study, \cite{drumm2020extensive}, mass effect was observed in 30\% of the cases involved in the study, and extensive GBM infiltration of the brainstem was present in 67\% of the GBM cases. Hence, it may be reasonable to assume that there might be latent disease-specific information captured via the subtle tissue deformations in the seemingly-normal brain regions adjacent to tumor (also known as ``brain around tumor (BAT)''), that may provide prognostic cues regarding GBM tumors. The rationale being
that more aggressive tumors may exert increased intra-cranial pressure on the surrounding BAT regions, resulting in pronounced structural deformations and
thus worse prognosis, as compared to less aggressive tumors.
 
Recently, ‘radiomics’ (i.e. extraction of quantitative image features such as co-occurrence, gray-level dependence, directional gradients, and shape-based) as well as deep-learning approaches using routine  magnetic resonance imaging (MRI) sequences have shown potential in survival prediction and response assessment for brain tumors \cite{lambin2012radiomics,kickingereder2016radiomic,yang2015evaluation,lao2017deep,chato2017machine,nie20163d}. These studies have focused on capturing local textural changes within the tumor and the peri-tumoral regions, and their associations with patient survival \cite{prasanna2017radiomic,beig2019perinodular}. However, a missing gap in previous work has been to leverage the subtle tumor-induced deformations in the BAT regions as measured on routine MRI scans, as a complementary radiomic marker for prognosticating patient survival. This combination of both the morphological patterns within the tumor confines as well as the biomechanical patterns of tumor deformations could allow for a comprehensive characterization of the disease and its field effect. 

In this work, we attempt to build an integrated MRI-based descriptor, that captures radiomic-Deformation and Textural Heterogeneity (r-DepTH) to comprehensively characterize tumor field effect, by leveraging both morphological and biomechanical deformation attributes of the tumor. The integrated r-DepTH descriptor comprises: (1) the subtle tissue deformations measured from within the BAT region, as a function  of the  distance  from  the  tumor  margins, and (2) textural radiomic features from the  visible tumor margins and peritumoral regions. The r-DepTH descriptor in our work is evaluated in the context of risk stratification for prognosticating overall patient survival in GBM tumors on $n = 207$ retrospective 1.5 Tesla (T) MRI studies (Gadolinium (Gd)-contrast $T_{1w}$, $T_{2w}$, $T_{2w}$-FLAIR), obtained from a multi-institutional cohort. Additionally, we compare r-DepTH with existing state-of-the-art radiomics \cite{davatzikos2018cancer} and deep-learning \cite{lao2017deep} based approaches, that have previously employed MRI-features from the tumor and peri-tumoral regions for GBM prognosis. 

This paper is organized as follows. In Section 2, previous work on characterizing field effect in survival prediction of GBM tumors using routine MRI scans is discussed. In Section 3, we describe the methodology for computation of the r-DepTH descriptor. The experimental design and implementation details for the risk assessment model are provided in Section 4, followed by results in Section 5. Section 6 provides discussion and concluding remarks. 

\section{Previous Work and Novel Contributions}
The concept of interrogating the tumor field using routine MRI scans has gained significant interest over the years, both to study its impact on tumor growth as well as correlating its impact on overall patient survival \cite{swanson2003virtual,harpold2007evolution,hormuth2017mechanically}. For instance,  works have previously developed deterministic mathematical models that model cancerous growths from aggressive cellular proliferation in GBM tumors \cite{swanson2003virtual}. 
Through these models, studies investigated the induced significant mechanical stress on the surrounding tissue that results in mass effect in GBM tumors \cite{swanson2003virtual,harpold2007evolution}. These mathematical models consider the cellular motility factor in GBMs, to account for its invasiveness as well as the ability of its cells to migrate and proliferate \cite{swanson2003virtual}. 
A mechanically coupled model was also suggested in \cite{hormuth2017mechanically}, to address the mechanical stress caused by tumor expansion, while also incorporating a diffusion coefficient that accounts for local tissue stress due to the field effect.  
In addition to these deterministic mathematical models, multiple studies have explored the utility of 'data driven' approaches, such as radiomic features extracted from GBM patients in survival prediction. For instance, \cite{kickingereder2016radiomic} showed that radiomic features outperformed clinical and radiologic risk models in predicting overall survival in GBM tumors. Similarly, the studies conducted by \cite{yang2015evaluation, gevaert2014glioblastoma, cui2015prognostic, mcgarry2016magnetic} assessed the utility of texture features extracted from the tumor and peritumoral regions for survival prediction in GBM. 
Our own group has developed a textural radiomic descriptor, Co-occurrence of Local Anisotropic Gradient Orientations (COLLAGE) \cite{prasanna2016co}, that has demonstrated success in capturing subtle differences between similar appearing disease phenotypes on routine imaging, across different types of malignant tumors \cite{braman2017intratumoral,shiradkar2018radiomic}. COLLAGE captures local anisotropic differences in intra-tumoral heterogeneity by calculating per-voxel gradient orientations, followed by obtaining statistics of Gray-Level Co-Occurrence Matrix (GLCM) heterogeneity to quantify patterns of local gradient alignment. However, those existing prognostic studies in GBM have been limited to interrogating texture representations from the enhancing lesion, inner necrotic core, and peritumoral area, and have not explicitly accounted for any possible biomechanical deformational changes in the BAT region. 

\begin{figure}[ht!]
   \begin{center}
  \includegraphics[width=3.5in]{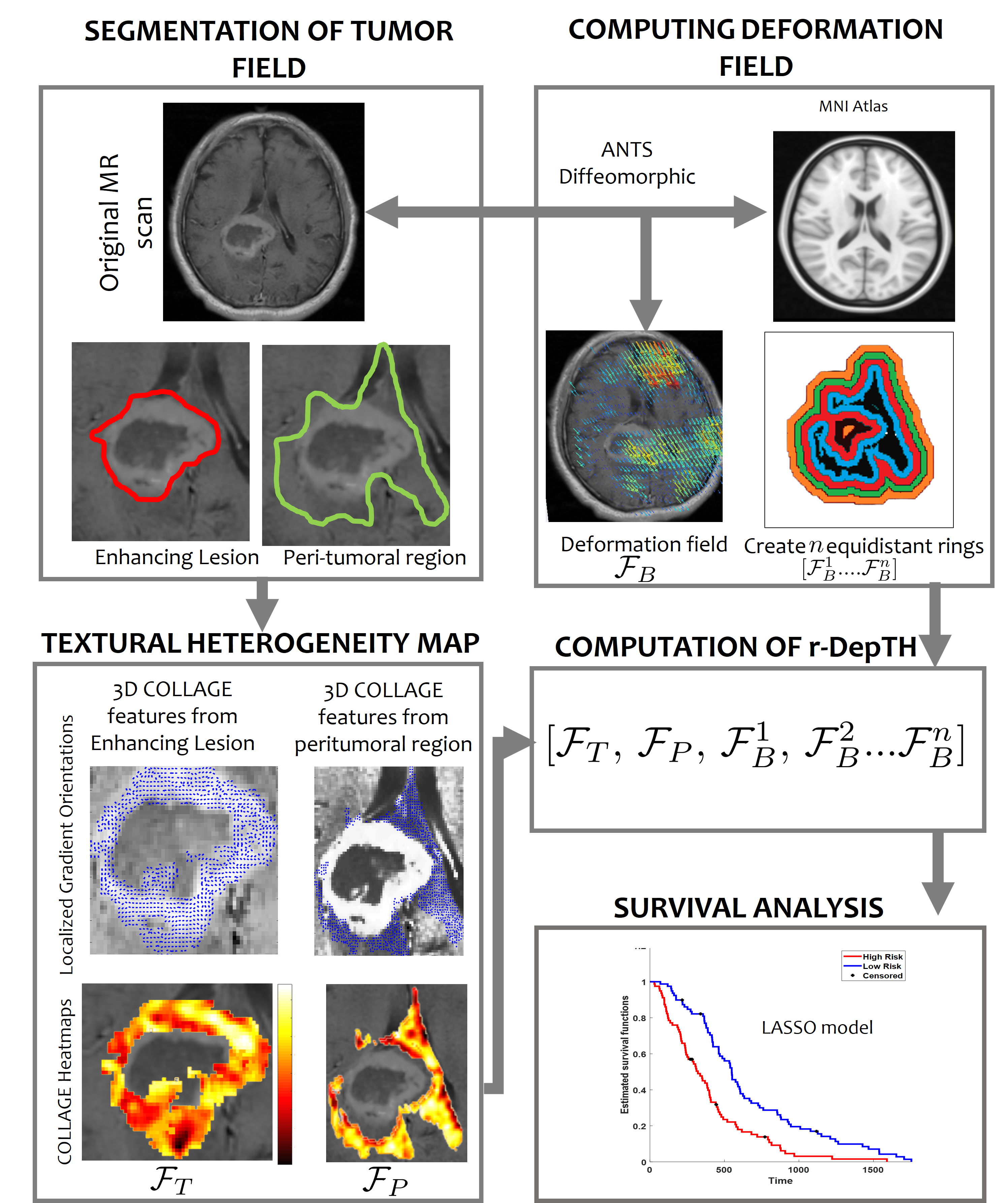}
   \end{center}
   \caption[Overview of rDepTH framework] 
Overview of r-DepTH framework. First, segmentation of tumor compartments of interest (enhancing lesion (outlined in red) and peri-tumoral area (outlined in green)) is performed. Following pre-processing, feature extraction is performed via COLLAGE features to characterize the intra-tumoral textural heterogeneity, and deformation heterogeneity features to characterize the tumor impact on BAT region. Next, the sets of COLLAGE and deformation features are concatenated to compute the r-DepTH descriptor. The r-DepTH descriptor could then be employed for classification/survival analysis (in our case using a LASSO model for stratifying GBM patients into low- and high-risk groups based on their overall survival.)\label{fig:framework}
   \end{figure}

The key contribution of this work is the development of a r-DepTH descriptor that combines measurements of biomechanical and morphological features of the tumor regions, for predicting patient survival. Briefly, we capture the biomechanical deformations in the BAT regions by using diffeomorphic registration between the GBM brain scans and a healthy atlas. We then utilize the inverse mapping during registration to calculate the deformation magnitudes within uniformly sized annular sub-volumes that belong to the surrounding BAT regions, both adjacent and distal to the tumor boundaries (as close as $5 mm$ and up to $60 mm$). In addition, we use 3D COLLAGE features \cite{prasanna2016co} to capture the textural heterogeneity from within the tumor visible confines. The deformation features from the normal parenchyma and the COLLAGE features from the tumor confines, are finally combined to obtain the integrated r-DepTH descriptor. Figure \ref{fig:framework} provides an overview of the r-DepTH framework.

\section{Methodology}
\subsection{Notation}
We define an image scene $I$ as $I = (C,f)$, where $I$ is a spatial grid of voxels $c \in C$, in a 3-dimensional space, $R^{3}$. Each voxel, $c \in C$, is associated with an intensity value $f(c)$. $I_T,I_P$, and $I_{B}$  correspond to the intra-tumoral, peri-tumoral, and surrounding normal parenchemal sub-volumes within every $I$ respectively, such that $[I_T,I_P, I_{B}] \subset I$. We further divide the sub-volume $I_B$ into uniformly sized annular sub-volumes $I_{B_{j}}$, where $j$ is the number of uniformly-sized annular bands, such that $j \in\{1,...,m\}$, and $m$ is a user-defined proximity parameter that is dependent on the distance from the tumor margin.

\subsection{R-DepTH Descriptor}

\subsubsection{Deformation heterogeneity features from the normal parenchyma}
Healthy T1-w MNI atlas $(I_{Atlas})$ is used to measure the tissue deformation in the normal appearing brain regions of every patient volume $I$. $I_{Atlas}$ is first non-rigidly aligned to $I$ using mutual-information-based similarity measure provided in ANTs (Advanced Normalization Tools) SyN (Symmetric Normalization) toolbox \cite{avants2008symmetric}. This toolbox is specifically used due to its proved efficiency in mapping brain images containing lesions into healthy templates \cite{eloyan2014health}. It also efficiently
handles the constrained cost-function masking approach, where the mapping within a tumor exclusive region is determined by the solution of the negative tumor mask region $I_{mask}$. We employ Lagrangian diffeomorphic-based registration \cite{avants2006landmark}, as it possesses symmetry properties required for a geodesic connecting two images in the space
of diffeomorphic transformation that guarantees symmetry regardless of the chosen similarity measure \cite{avants2008symmetric}. We also wanted to ensure that only the intensity differences due to structural deformations are accounted for, during registration, while excluding the intensity differences within the tumor area. $I_{mask}$ was hence removed from $I$ during registration to $I_{Atlas}$. Given the reference $(I)$ and floating $(I_{Atlas})$, the non-rigid alignment can be formulated as: $(I, I_{mask}) = Tr(I_{Atlas})$, where $Tr(.)$ is the forward transformation of the composite voxel-wise deformation field (including affine components) that maps the displacements of the voxels between the reference and floating volumes. This transformation also propagates the atlas brain mask $(I_{atlas})$ to the subject space, thereby skull stripping the subjects. As ANTs SyN satisfies the conditions of a diffeomorphic registration, an inverse $Tr^{-1} (.)$ exists, that successfully maps $I$ to the $I_{Atlas}$ space. This inverse mapping yields the tissue deformation of $I$ with respect to $I_{Atlas}$, representing the deformations exerted on every $c \in C_{B}$, due to tumor mass effect. Considering $(c' _t,c' _u,c' _v)$ as new voxel positions of $I$ when mapped to $I_{Atlas}$, the displacement vector is given as $[\delta t,\delta u,\delta v]$ where vector $(c' _t,c' _u,c' _v )= (c_t,c_u,c_v )+(\delta t,\delta u,\delta v)$, and the magnitude of deformation is given by calculating the Euclidean norm of the scalar values of the deformation orientations as:
\begin{equation}
D(c)= \sqrt{(\delta t)^2+(\delta u)^2+(\delta v)^2},
\end{equation}
for every $c \in C_{B_j}$, and $j \in \{1,.....,m\}$. First order statistics (i.e. mean, median, standard deviation, skewness, and kurtosis) are then calculated by aggregating $D(c)$ for every $c$ within every sub-volume $I_{B_j}$ yielding a feature descriptor $\mathcal{F}_{B_{j}}$ for every annular sub-region $C_{B_j}$, where $C_{B_j}  \subset C_B$, $j \in \{1,...,m\}$. 

\subsubsection{3D COLLAGE features from within the tumor confines}
 COLLAGE, a 3D gradient-based texture descriptor, captures intra-tumoral heterogeneity by calculating local per-voxel gradient orientations \cite{prasanna2016co}. Briefly, for every voxel $c$, intensity gradients in $X,Y,Z$ directions are calculated, followed by centering a $3D$ window around every $c$ to compute the vector gradient matrix $F$. Then, two principal orientations, $\theta(c)$ and $\phi(c)$, can be obtained from $F$ for every $c$, followed by computing two separate co-occurrence matrices, $M^\theta$ and $M^\phi$, that capture orientation pairs between voxels in a local neighborhood. From every co-occurrence matrix, a total of 13 Haralick statistics are calculated for every $c$ \cite{haralick1973textural}. We finally obtain first order statistics (mean, median, standard deviation, skewness, and kurtosis) for every $c \in \{C_P,C_T\}$, which yields a feature descriptor $\mathcal{F}_{T}$  for the enhancing lesion, and $\mathcal{F}_{P}$ for the T2/FLAIR hyperintensities. Additional details regarding COLLAGE computation can be found in \cite{prasanna2016co}.

\subsubsection{Construction of r-DepTH descriptor}
The r-DepTH descriptor is obtained for every patient, by concatenating the deformation descriptor, $\mathcal{F}_{B}$, along with COLLAGE texture descriptors, $\mathcal{F}_{T}$, and $\mathcal{F}_{P}$, as $\mathcal{F}_{rDepTH} = [\mathcal{F}_{B}, \mathcal{F}_{T}, \mathcal{F}_{P}]$. $\mathcal{F}_{rDepTH}$ descriptor can then be employed within a supervised or an unsupervised approach for disease characterization. The algorithm for computing r-DepTH is provided in Algorithm \ref{alg:rdepth}.

\begin{algorithm}
\footnotesize
\caption{Computation of r-DepTH descriptor}
\label{alg:rdepth}
\hrule height 0.1pt
\vspace{5pt}
\KwData{$I$, $I_{Atlas}$, ROI Volume $C$}
\KwResult{$\mathcal{F}_{rDepTH}$}
\vspace{4pt}
\hrule height 0.1pt
\vspace{4pt}
\SetAlgoNoLine
\Begin{
\textbf{1- Obtain Deformation Field $\mathcal{F}_{B}$}\\
 \For{each patient volume $I$} {
     Remove $I_{mask}$ from $I$, align  $I_{Atlas}$ to $I$ to get $(I,I_{mask})$ $=$ $Tr(I_{Atlas})$ \\
     }
     \For {each $c \in C_{B}$}{
      Get the deformation of $I$ w.r.t. $I_{Atlas}$, $[\delta t,\delta u,\delta v]$, using $Tr^{-1} (.)$.
      
     Get deformation magnitude $D(c)= \sqrt{(\delta t)^2+(\delta u)^2+(\delta v)^2}$. \\
     }
     \For {each annular sub-region $C_{B_j}  \subset C_B$}{
     Aggregate $D(c)$ for every $c$ within sub-volume $I_{B_j}$. \\
     Calculate first order statistics for $I_{B_j}$ to get $\mathcal{F}_{B_{j}}$. \\
     }
     
\textbf{2- Obtain 3D COLLAGE Features $\mathcal{F}_{T}, \mathcal{F}_{P}$}\\
 \For{each voxel $c \in C$} {
  Obtain gradients $\partial f_i(c)$ along $i$-axes, $\partial f_i(c)=\frac{\partial f(c)}{\partial i}$,  $i \in {x,y,z} $.
     }
     Define $\mathcal{N} \times \mathcal{N} \times \mathcal{N}$ neighborhood centered around $c \in C$. \\
     \For{each voxel $c \in C$}{
     Calculate gradient vectors $\overrightarrow{\partial f_i}(c_k)$ in $\mathcal{N}\times\mathcal{N}\times\mathcal{N}$, $i \in {(x,y,z)} $, $k \in \{1,\dots,\mathcal{N}^3\}$, where
 $\partial f_i(c)=\begin{bmatrix}
         {\partial f}_i(c_1)   &   {\partial f}_i(c_2)&..&   {\partial f}_i(c_{N^3})
        \end{bmatrix}^{T} $\\
 Obtain localized gradient vector matrix $\overrightarrow{\mathcal{F}}=[\overrightarrow{\partial f_X}(c_k)\;\;\overrightarrow{\partial f_Y}(c_k)\;\;\overrightarrow{\partial f_Z}(c_k)]$\\

Calculate dominant components $\psi(c_{k11})$, $\psi(c_{k12})$, and $\psi(c_{k13})$, $k \in \{1,\dots,\mathcal{N}^3\}$,  by singular value decomposition of $\overrightarrow{\mathcal{F}}$\\
Obtain dominant directions $\theta(c_k)$ and $\phi(c_k)$, using $\theta^{3D}(c_k) = \tan ^{- 1}\frac{\psi_Y(c_{k})}{{\psi_X(c_{k})}}$ and $\phi^{3D}(c_k) = \tan ^{- 1}\frac{\psi_Z(c_{k})}{\sqrt{\psi_Y^2(c_{k}) + \psi_X^2(c_{k}})}$ \\
 }
Compute co-occurrence matrices $M^\theta$ and $M^\phi$ from $\theta (c_k)$  and $\phi (c_k)$\\
      
     \For{each voxel $c \in \{C_P, C_T$\}}{
     Get 13 Haralick statistics $[S_{\theta_b} ,S_{\phi_b}]$, $b \in [1,13]$ from $M^\theta$ and $M^\phi$\\
     Calculate first order statistics to get $\mathcal{F}_{T}, \mathcal{F}_{P}$ 
     }
     
 \textbf{3- Compute r-DepTH $\mathcal{F}_{rDepTH}$}\\
 Concatenate $\mathcal{F}_{B}$, $\mathcal{F}_{T}$, and $\mathcal{F}_{P}$ to get $\mathcal{F}_{rDepTH}$
}
\hrule height 0.1pt
\end{algorithm}

\section{Experimental Design}
While the r-DepTH descriptor is generalizable, in this work, we calculated the $\mathcal{F}_{rDepTH}$ from routine pre-treatment MRI scans as a prognostic marker for overall survival in GBM subjects. Details on experimental design are provided below. 

\subsection{Data}
A total of $207$ cases were collected from multiple sites for this study, including the publicly available Cancer Imaging Archive (TCIA) (\cite{clark2013cancer}), Cleveland Clinic (CCF), and Medical College of Wisconsin (MCW). TCIA is an open archive of radiological scans for different cancer types including GBM consisting of MRIs and its associated clinical metadata, with regulations and policies for the protection of human subjects and approvals by institutional review boards in place. Data analysis from Cleveland Clinic and MCW was approved by the institutional Ethics Committee. Our inclusion criteria included the availability of (1) 1.5 Tesla (T) routine MRI sequences (Gadolinium (Gd)-contrast $T_{1w}$, $T_{2w}$, $T_{2w}$-FLAIR) for treatment-naive patients with diagnostic image quality, and (2) Overall Survival (OS) information. This inclusion criteria yielded a total of (1) $75$ GBM studies from TCIA, (2) $53$ studies from CCF, and (3) $79$ studies from MCW. The cases from the TCIA and CCF were combined for training ($S_{t}$), whereas the cases from the MCW were held-out for testing ($S_{v}$). The OS for the $128$ cases used for $S_{t}$ had a mean of $509.07$ days, whereas the mean of OS for $S_{v}$ was $468.7$ days, as detailed in Table \ref{tab:data}. Section 1 in the supplementary material provides additional details regarding the data employed in our work.


\begin{table}
\centering       
\begin{tabular}{|P{2.4cm}|P{3.7cm}|P{1.4cm}|P|} 
\hline
Group & $S_{t}$ &  $S_{v}$\\ 
\hline
Number  & 128 &79\\
\hline
Mean of age (Years) & 59.4 &  61.5\\
\hline 
Gender & 80 males, 48 females &  \begin{tabular}{c}39 males\\ 40 females \end{tabular}\\
\hline
OS (mean $\pm$ STD) (days) & 509.07 $\pm$ 379 &  468.7$\pm$430.8\\
\hline
Censored Subjects & 3 &  0\\
\hline
\begin{tabular}{c}Extent of Resection\\ (EOR) \end{tabular} & \begin{tabular}{c}n = 55 \\  subtotal resection (STR): 23 \\ gross total resection (GTR): 16\\ Neuroblate: 7\\ near total resection (NTR): 9 \end{tabular}  & NA\\
\hline
MGMT &\begin{tabular}{c}n = 99\\Methylated: 46 \\ Non-methylated: 53 \end{tabular}&NA\\
\hline
IDH &\begin{tabular}{c}n = 128\\Wild Type: 120 \\ Mutant: 8 \end{tabular}&NA\\
\hline
\end{tabular}
\caption{Data description for both $S_{t}$ and $S_{v}$} 
\label{tab:data}
\end{table} 

\subsection{Preprocessing}
A total of three experts were asked to perform the manual annotations on every MRI slice, via a hand-annotation tool in 3D Slicer. The senior-most expert (V.H, $>$10-years of experience in neuro-radiology) independently annotated the studies obtained from CCF, while expert 2 (V.S) with 7 years of experience in neuro-radiology supervised expert 3 (K.B, with 3 years of radiology experience), to manually annotate the TCIA and MCW cases. In rare cases with disagreement across the two readers (expert 2 and expert 3), the senior-most radiologist (V.H, expert 1) was consulted to obtain the final segmentations.  Every tumor was annotated into 2 regions: enhancing lesion ($I_T$) and $T_{2w}$/FLAIR hyperintense peri-lesional component ($I_P$). $T_{1w}$ MRI scans were used to delineate $I_T$, while both $T_{2w}$ and FLAIR scans were used to annotate $I_P$. Following segmentation, for every patient study, the 3 MRI sequences, Gd-$T_{1w}$ MRI, $T_{2w}$, and FLAIR were co-registered to a brain atlas (MNI152; Montreal Neurological Institute) using ANTs (Advanced Normalization Tools) SyN (Symmetric Normalization) toolbox \cite{avants2008symmetric}. Skull stripping was performed simultaneously during registration of $I$ with $I_{Atlas}$, as detailed in Section 3.2.1. Finally, bias field correction was conducted using a non-parametric non-uniform intensity normalization technique \cite{tustison2010n4itk}. 
\subsection{Implementation Details}
We calculated deformation magnitudes $\mathcal{F}_{B_j}$ for $j \in \{1,2, \dots, 12\}$ annular regions that are equidistant to each other at a distance of $5mm$. The choice for the size of the annular rings is based on retrospective studies that have recommended $5mm$ as safe clinical target volume margin for GBM tumors \cite{cabrera2016radiation}. Creation of annular bands also ensured consistency in our analysis as well as sufficient number of voxels for statistics computations. This resulted in a $60 \times 1$ deformation vector that included 5 statistics (mean, median, standard deviation, skewness, and kurtosis) calculated for each of the 12 bands. This resulted in $12 \times 5 = 60$ features corresponding to $\mathcal{F}_{B}$. In addition, the 5 statistics were similarly obtained for each of the 13 Haralick statistics across each of the two co-occurrence matrices, resulting in  $13 \times 5 \times 2 = 130$ COLLAGE features that are extracted from each of the two compartments; $\mathcal{F}_{T}$, and $\mathcal{F}_{P}$.
The descriptor $\mathcal{F}_{rDepTH}$ was finally obtained by aggregating $\mathcal{F}_{B}$, $\mathcal{F}_{T}$, and $\mathcal{F}_{P}$. Hence the  $\mathcal{F}_{rDepTH}$ descriptor for every tumor included a total of $320$ features. 

\subsection{Survival risk assessment}
 Following computation of $\mathcal{F}_{\alpha}$, where $\alpha=\{T,P,B,rDepTH\}$, feature selection (reduction) was conducted using least absolute shrinkage and selection operator (LASSO), along with a cox regression model \cite{tibshirani1996regression} on $S_{t}$.   
We used LASSO to utilize its capability in reducing variance when shrinking features, while simultaneously not increasing the bias. The top features selected by LASSO model were then used to create a continuous survival risk score ($Risc$), calculated as:
\begin{equation}
    Risc({\mathcal{F}_{\alpha}}) = \sum_{g=1}^{A}q_g\mathcal{F}_{\alpha}^g
\end{equation}

Where $A$ represents the number of selected imaging features from LASSO, $\mathcal{F}_{\alpha}^g$ is the $g$th feature for $\alpha=\{T,P,B,rDepTH\}$ and $q_g$ is the respective coefficient. 
An observation was deemed censored if the patient withdrew from the study or there was no follow up available. We used X-tile software (version 3.6.1) to find an optimal threshold using a grid-search across all $Risc$ values to classify patients into high-risk (H-R) and low-risk (L-R) groups obtained from $S_{t}$. Log-rank test along with Kaplan-Meier (KM) survival analysis were then performed to see how the survival rate varies between the identified L-R and H-R groups. Additionally, performance measures such as hazard ratios (HR), 95\% Confidence Interval (CI), and Concordance index (C-index) were obtained to assess the performance of our survival models. HR is defined as the risk of experiencing the event of interest at a time point \cite{austin2017median}, whereas CI measures the level of uncertainty about the point estimates \cite{pencina2004overall}. C-index was calculated using R (v4.0), and is commonly used to validate the predictive ability of a survival model by calculating the probability of concordance between the predicted and the observed survival \cite{steck2008ranking}. Finally, on $S_{v}$, the top features obtained from $S_{t}$ were used to calculate $Risc$ for every patient, followed by the log-rank test to obtain the level of significance between the two identified groups (L-R and H-R).

\subsection{Comparative Strategies}
In order to evaluate the efficacy of r-DepTH for GBM survival prediction, we performed the following comparisons: (1) Employing clinical features (age, gender, tumor volume, molecular information), in uni-variate and multi-variate settings, 
(2) Evaluating $Risc({\mathcal{F}_{T,P}})$ using $\mathcal{F}_{T}$ and $\mathcal{F}_{P}$, along with age and gender, (3) Evaluating $Risc({\mathcal{F}_{B}})$ using $\mathcal{F}_{B}$, along with age and gender,  (4) State-of-the-art radiomics \cite{davatzikos2018cancer}, and convolutional neural network (CNN)  \cite{lao2017deep} approaches, previously used in the literature for GBM prognosis. 

Table \ref{tab:comp} details the feature families as well as the number of features extracted in the comparative radiomic approach, from $[C_T, C_P]$ regions, using the CapTk software as described in \cite{davatzikos2018cancer}. This resulted in a total of 4376 features ($\mathcal{F}_{Rad}$) for every tumor region in $S_{t}$. This was followed by feeding ($\mathcal{F}_{Rad}$), along with age and gender information, to our LASSO model to calculate $Risc({\mathcal{F}_{rad}})$ for every patient. 

Additionally, we compared the performance of the r-DepTH descriptor to a CNN model previously utilized in the literature \cite{lao2017deep}, to predict survival in GBM. This model is similar to most of the models found in literature in the context of survival prediction in GBM, where transfer learning has been exploited to stratify patients into risk groups. Specifically, we extracted deep features from the GBM patients using a pre-trained CNN model via transfer learning, along with radiomic features, to predict survival. The network contained 5 convolutional layers and 3 fully-connected layers. The model was trained on ImageNet database with predetermined weights that are summarized in Table \ref{tab:comp}. The input to the model was the cropped tumor sub-regions from the MR scans, obtained from the slice that had the largest tumor area for every patient, followed by resizing the sub-regions to $224 \times 224$ with bicubic interpolation. Deep features were computed by forward propagation, and extracted from the fully-connected layer 7. A total of 4096 features ($\mathcal{F}_{DL}$) were extracted from each patient, which were finally fed to our LASSO model for survival prognostication, to compute
$Risc({\mathcal{F}_{DL}})$ for every patient. All the comparative strategies conducted in this work are summarized in Table \ref{tab:comp}.



\begin{table}[ht!]

\centering       
\begin{tabular}{ |c|c|P{4.3cm}|c| } 
\hline
Approach & No. features & \begin{tabular}{c} Features/\\Parameters \end{tabular}\\ 
\hline
Radiomic-based   & 4376 & \begin{tabular}{c} maximum, minimum,\\ variance, standard deviation,\\ skewness, kurtosis,\\ mean, and median of: \\Intensity histogram distributions,  \\ Texture, Shape, and\\ Spatial features \end{tabular}\\
\hline
CNN-based & 4096 &  \begin{tabular}{c}\begin{tabular}{c} Weight decay = $5 \times 10^{-4}$  \end{tabular} \\ \begin{tabular}{c} Momentum = $0.9$  \end{tabular} \\ \begin{tabular}{c} Initial learning rate = $10^{-2}$  \end{tabular} \end{tabular}\\
\hline
Clinical & 6 &  \begin{tabular}{c}\begin{tabular}{c} Uni- or Multi-variate settings of:  \end{tabular} \\ \begin{tabular}{c} Age, Gender, \end{tabular} \\ \begin{tabular}{c} Tumor Volume, EOR,  \end{tabular}  \\ \begin{tabular}{c} MGMT, IDH  \end{tabular} \end{tabular}\\
\hline
COLLAGE & 260 &  \begin{tabular}{c}\begin{tabular}{c} Mean, Median, Skewness,  \end{tabular} \\ \begin{tabular}{c} Standard deviation, Kurtosis \end{tabular} \\ \begin{tabular}{c} of the 13 Haralick statistics   \end{tabular}  \\ \begin{tabular}{c} for each tumor region  \end{tabular} \end{tabular}\\
\hline
Deformation & 60 &  \begin{tabular}{c}\begin{tabular}{c} Mean, Median, Skewness,  \end{tabular} \\ \begin{tabular}{c} Standard deviation, Kurtosis \end{tabular} \\ \begin{tabular}{c} of the 12 annular bands   \end{tabular}  \\ \begin{tabular}{c} of the BAT region \end{tabular} \end{tabular}\\
\hline
\end{tabular}
\caption{Comparative strategies to r-DepTH} 
\label{tab:comp}
\end{table} 

\section{Results}

\subsection{Survival risk assessment using r-DepTH}
LASSO survival analysis while creating $Risc({\mathcal{F}_{rDepTH}})$ yielded $9$ features (listed in supplementary material). The KM curve obtained for $S_{t}$ based on these 9 features (Figure \ref{fig:KM}(e)) demonstrated significant differences between the two groups, $p$-value = $3.5 \times 10^{-7}$. The C-index obtained using $\mathcal{F}_{rDepTH}$ was $0.8$, with HR (CI) = $2.2$  $(1.5 - 3)$.
Similarly, the KM curve obtained for $S_{v}$ using the top 9 features demonstrated  significant differences between the two L-R and H-R survival groups (Figure \ref{fig:KM}(j)), with $p$-value = $0.0024$, C-index = $0.75$, and HR (CI) = $1.94$  $(1 - 3)$.

\begin{figure*}[ht!]
   \begin{center}
   \includegraphics[height=9cm, width = 17cm]{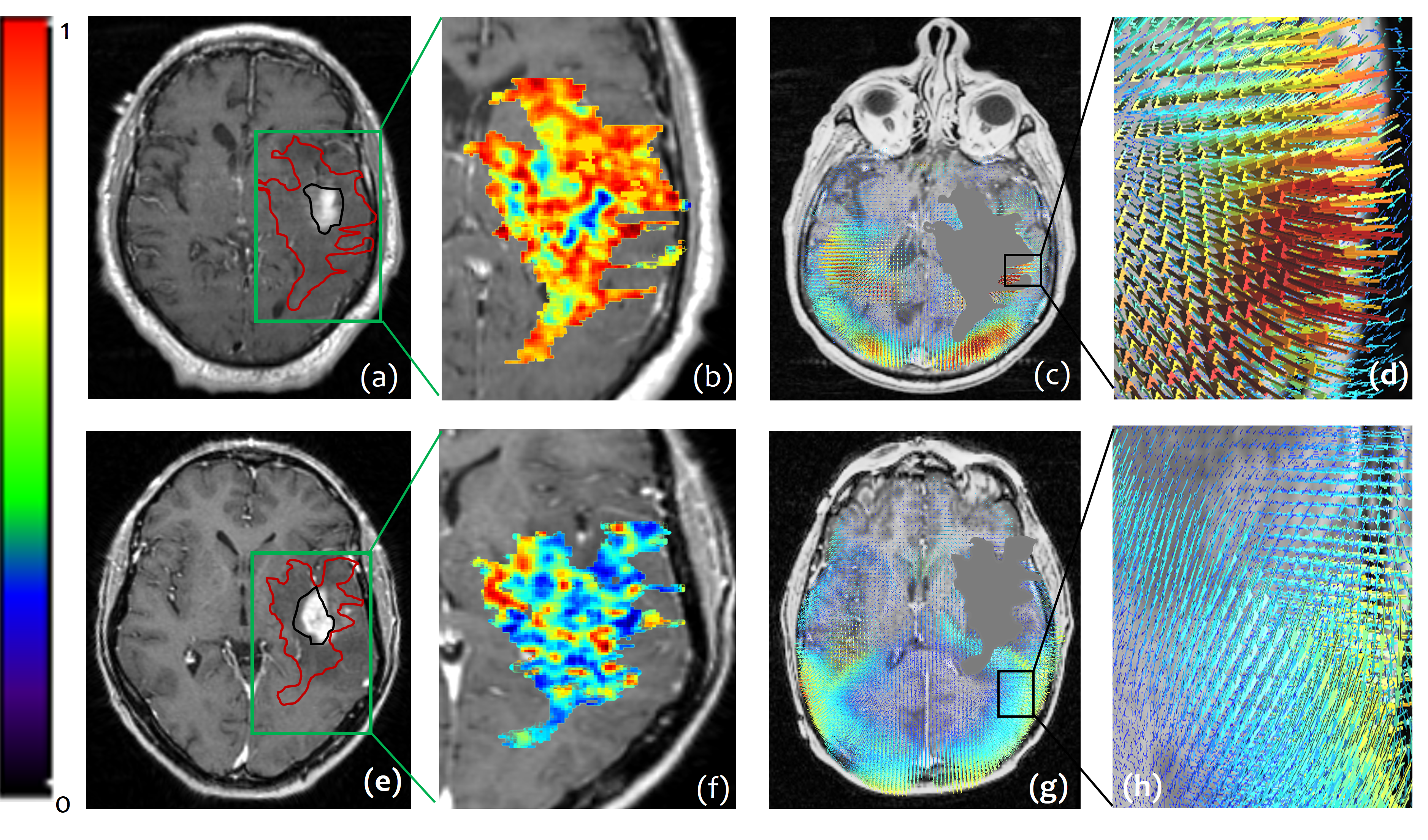}
   \end{center}
   \caption[TCIA1] 
   { \label{fig:res} 
Two subjects from the $S_{t}$; a patient with poor survival (top row), OS $=$ 30 days, and a patient with prolonged survival (bottom row), OS $=$ 691 days. (a), (e) show the corresponding Gd-T1w MR scans of the two patients with their tumors segmented into 2 compartments: enhancing lesion (outlined in black) and peri-tumoral area (outlined in red). (b), (f) demonstrate the COLLAGE heatmaps generated for each of the two subjects, with higher values (red) being more prevalent in the patient with poor survival, compared to the patient with prolonged survival. (c), (g) illustrate the extracted deformation field magnitudes respectively for each of the two patients. For the patient with poor survival (d),  higher magnitude values (represented in red) were observed in close proximity of the tumor, whereas lower values were  observed (blue) for the patient with prolonged survival (h).} 
   \end{figure*}

Qualitative differences for both COLLAGE and deformation fields for two GBM subjects, one with poor-survival (OS = 30 days) (top row) and one with prolonged-survival (OS = 691 days) (bottom row), are presented in Figure \ref{fig:res}. The patient with poor-survival seemed to exhibit higher values of the COLLAGE feature (Kurtosis of Energy) (Figure \ref{fig:res}(b)), 
compared to the patient with prolonged survival (Figure \ref{fig:res}(f)). Similarly, the deformation field magnitudes for Skewness (measure of data asymmetry) at $10mm$ are visualized, and seem to reflect higher values for the patient with worse survival (Figure \ref{fig:res}(d)), both in close proximity as well as distal to the tumor, compared to that for the patient with improved survival (Figure \ref{fig:res}(h)). In addition, box-plots for the 3 most discriminative COLLAGE and deformation features on $S_{t}$ and $S_{v}$ are shown in Figure \ref{fig:box}.
The top deformation features included skewness and kurtosis of $\mathcal{F}_{B_5}$ as well as skewness of $\mathcal{F}_{B_{10}}$. Similarly, the $3$ top COLLAGE features were the skewness of sum variance as well as kurtosis of both energy and sum average.


\subsection{Comparative approaches}
\subsubsection{Risk-scores using  clinical information}
 Age alone demonstrated significant differences in the survival groups on $S_{t}$, but the differences were not significantly different on $S_{v}$. Gender and tumor location, both in a univariate and multi-variate setting, did not yield significant differences between the two risk-groups (L-R versus H-R) across $S_{t}$ and $S_{v}$, as shown on Table \ref{tab:stats}. 
Additionally on $S_{t}$, we evaluated molecular markers including MGMT (available for 99 subjects), and IDH (available for 128 subjects), as well as extension of resection (EOR) (available for 55 subjects) for prognosis of GBM survival, and did not observe any significant differences between the two survival groups. These clinical and molecular parameters unfortunately were not available for $S_{v}$ and hence could not be evaluated.  $p$-values, C-index and HR (CI) for each of these experiments are provided in Table \ref{tab:stats}.

\subsubsection{Survival risk-assessment using COLLAGE features from the tumor and peri-tumoral regions}
$Risc({\mathcal{F}_{T,P}})$ constituted of a total of 10 features, 8 from $\mathcal{F}_{T}$ and 1 from $\mathcal{F}_{P}$ in addition to the age attribute, obtained from the LASSO model. Some of these top features included standard deviation of Correlation for $\mathcal{F}_{T}$, and skewness of Sum Variance (for both $\mathcal{F}_{T}$ and $\mathcal{F}_{P}$). A complete list of all 10 features is provided in supplementary material. The KM curve (Figure \ref{fig:KM}(c)) obtained  from $Risc({\mathcal{F}_{T,P}})$ demonstrated  significant differences between the L-R and H-R survival groups, $p$-value = $4 \times 10^{-5}$. The C-index obtained using features that formed $Risc({\mathcal{F}_{T,P}})$ was $0.75$, with HR (CI) = $1.9$  $(1.4 - 2.7)$, on $S_{t}$.
KM curve on $S_{v}$ demonstrated  significant differences between the two L-R and H-R survival groups (Figure \ref{fig:KM}(h)) with $p$-value = $0.0073$, C-index of $0.65$, and HR (CI) of $1.5$  $(0.9 - 2.7)$ respectively.

\subsubsection{Survival risk assessment using deformation features from the tumor and peri-tumoral regions}
The LASSO survival analysis to obtain  $Risc({\mathcal{F}_{B}})$ yielded a total of 10 features, 2 features from $\mathcal{F}_{B_5}$, 1 from $\mathcal{F}_{B_{20}}$, 1 from $\mathcal{F}_{B_{25}}$, 1 from $\mathcal{F}_{B_{35}}$, 2 from $\mathcal{F}_{B_{40}}$, and 1 from $\mathcal{F}_{B_{60}}$, in addition to age and gender attributes (details are in supplementary material). The KM curve obtained for $S_{t}$ using $Risc({\mathcal{F}_{B}})$ (Figure \ref{fig:KM}(d)) demonstrated significant differences between the two groups, $p$-value = $2 \times 10^{-6}$. The C-index obtained using the features that yielded $Risc({\mathcal{F}_{B}})$ was $0.73$, with HR (CI) = $2$  $(1.5 - 3)$, on $S_{t}$.
Similarly, the KM curve obtained for $S_{v}$ for $Risc({\mathcal{F}_{B}})$ demonstrated significant differences between the two L-R and H-R groups (Figure \ref{fig:KM}(i)), with $p$-value = $0.004$, C-index of $0.68$, and HR (CI) = $1.5$  $(0.9 - 2.5)$. 

\subsubsection{Survival risk assessment using comparative radiomic and CNN approaches}
\begin{figure*}[ht!]
   \begin{center}
   \includegraphics[height=8.5cm, width = 17.5cm]{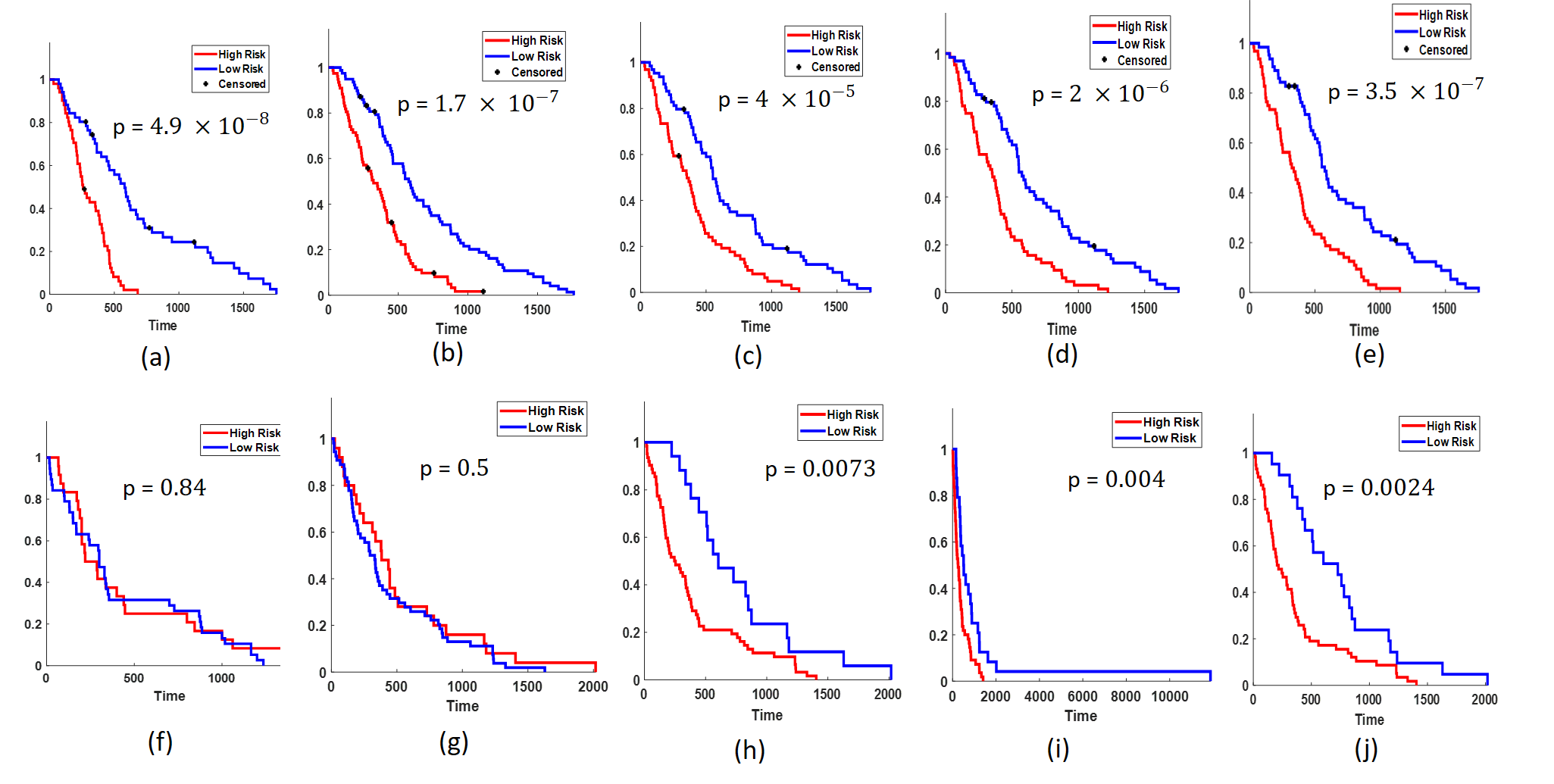}
   \end{center}
   \caption[TCIA1] 
   { \label{fig:KM} 
Kaplan Meier curves for estimating overall survival for our GBM cohort on $S_{t}$ (Top) and $S_{v}$ (Bottom) using (1) the comparative radiomic approach $Risc({\mathcal{F}_{Rad}})$ (a), (f), (2) the DL approach $Risc({\mathcal{F}_{DL}})$ (b), (g), (3) $Risc({\mathcal{F}_{T,P}})$  (c), (h), (4) $Risc({\mathcal{F}_{B}})$ (d), (i), and (5) $Risc({\mathcal{F}_{rDepTH}})$ (e), (j). X-axis represents the overall survival days in days, and Y-axis represents the estimated survival function.} 
   \end{figure*}

$Risc({\mathcal{F}_{Rad}})$ comprised of a total of 20 features (as listed in supplementary material) obtained from the LASSO model. While these features showed significant differences between L-R and H-R survival groups on $S_{t}$ (Figure \ref{fig:KM}(a)), with $p$-value = $4.9 \times 10^{-8}$, C-index = $0.75$, and HR (CI) = $1.9$  $(1.4 - 2.4)$, this feature set did not demonstrate significant differences on $S_{v}$ (Figure \ref{fig:KM}(f)) with $p$-value = $0.84$, C-index = $0.6$, and HR (CI) = $0.9$  $(0.54 - 1.54)$.
In contrast, the LASSO analysis to obtain $Risc({\mathcal{F}_{DL}})$ yielded a total of 35 features. These features, on $S_{t}$, showed significant differences between L-R and H-R survival groups as reflected in the KM curve (Figure \ref{fig:KM}(b)), with $p$-value = $1.7 \times 10^{-7}$, C-index = $0.8$, and HR (CI) = $2$  $(1.8 - 3)$. On $S_{v}$, there were no significant differences across the two groups (Figure \ref{fig:KM}(g)), with  $p$-value = $0.54$, C-index = $0.5$, and HR (CI) = $1.1$  $(0.7 - 1.7)$.

   \begin{table*}
 \centering
 \begin{tabular}{|c|c|c|c|c|c|c|c|}
 \hline
 & \multicolumn{1}{c|}{Feature} & \multicolumn{3}{c|}{$S_{t}$} & \multicolumn{3}{c|}{$S_{v}$}\\ 
 \cline{3-8}
    &  & $p$-value & C-index & HR (95\% CI)  & $p$-value  & C-index & HR (95\% CI)\\
 \hline
 \parbox[t]{2mm}{\multirow{6}{*}{\rotatebox[origin=c]{90}{Univariat}}} & Age &0.045&0.58&1.4 (1.02 - 2.1)& 0.21& 0.59& 1.36 (0.86 - 2.15)\\   \cline{2-8}
 & Gender &0.58&0.53&1.1 (0.78 - 1.6)&0.32&0.53& 0.77 (0.5 - 1.2)\\   \cline{2-8}
 & EOR &0.13&0.6&1.3 (1.2 - 1.5)&-&-&-\\  \cline{2-8}
 & IDH &0.5&0.52&1.1 (0.8 - 1.2)&-&-&-\\  \cline{2-8}
 & MGMT &0.3&0.52&1 (0.7 - 1.3)&-&-&-\\  \cline{2-8}
 & Tumor Volume &0.05&0.58&1.4(1.01 - 1.95)&0.59 & 0.46& 0.86 (0.54 - 1.35)\\  \hline
 \hline  
 \parbox[t]{2mm}{\multirow{5}{*}{\rotatebox[origin=c]{90}{Multivariate}}}  & Age + Gender &0.29 & 0.57 &1.2 (0.87- 1.67) &0.28 & 0.59 & 1.3 (0.84 - 2.1)\\  \cline{2-8}
 & IDH + MGMT & 0.23 & 0.59 & 1.25 (0.89 - 1.7)& - & - & - \\ \cline{2-8}
 & COLLAGE & 0.0004 & 0.75 & 1.9 (1.4 -2.7) & 0.0073 & 0.65& 1.5 (0.9 - 2.7)\\  \cline{2-8}
 & Deformation & 0.000002 & 0.73 & 2 (1.5 -3) & 0.004 & 0.68 & 1.5 (0.9 - 2.5)\\  \cline{2-8}
 & Radiomics &0.00000005& 0.75 & 1.9 (1.4 - 2.4) & 0.84 & 0.6 & 0.9 (0.54 - 1.54) \\ \cline{2-8}
  & DL &0.00000017& 0.8 & 2 (1.8 - 3) & 0.54 & 0.5 & 1.1 (0.7 - 1.7) \\ \cline{2-8}
  & \textbf{rDepTH} & \textbf{0.00000035}& \textbf{0.8} & \textbf{2.2 (1.5 - 3)} & \textbf{0.0024} & \textbf{0.75} & \textbf{1.94 (1 - 3)} \\ 
 \hline

 \end{tabular}
  \caption{$p$-value, concordance index, hazard ratio, and 95\% confidence interval for the experiments conducted, on $S_{t}$ and $S_{v}$ respectively.} 
  \label{tab:stats}
 \end{table*}

\begin{figure*}[ht!]
   \begin{center}
   \includegraphics[height = 6.5cm, width = 17cm]{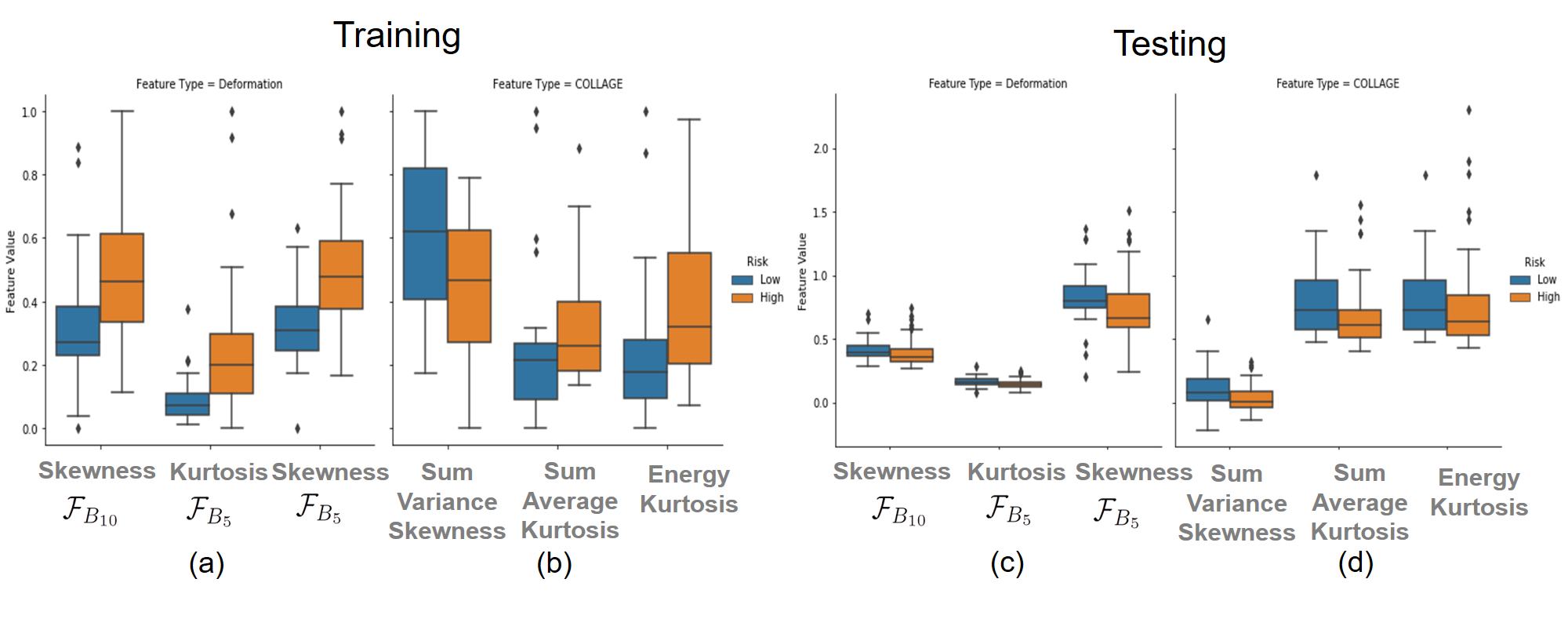}
   \end{center}
   \caption[plots] 
   {\label{fig:box}
Box plots of the 3 most significantly different deformation (a), (c) and COLLAGE (b), (d) features between patients with poor OS and those with prolonged OS, of the $S_{t}$ and $S_{v}$ respectively, based on the LASSO regression model. $p$-values for the differences in deformation features across the 2 risk groups (Skewness ($\mathcal{F}_{B_{10}}$), Kurtosis ($\mathcal{F}_{B_5}$), and Skewness ($\mathcal{F}_{B_5}$)) were $0.002$, $3 \times 10^{-4}$, and $3 \times 10^{-4}$ respectively for $S_{t}$, and $0.03$, $0.04$, and $0.02$ respectively for $S_{v}$. $p$-values for the differences in COLLAGE features (Sum Variance Skewness, Sum Average Kurtosis, and Energy Kurtosis) across the 2 groups were $0.015$, $0.03$, and $0.002$ respectively for $S_{t}$, and $0.04$, $0.03$, and $0.05$ respectively for $S_{v}$.}
   \end{figure*}
   
\section{Discussion}
Highly aggressive tumors such as Glioblastoma (GBM) tend to proliferate way beyond their visible tumor confines on routine MRI scans. For instance, GBM tumors are known to displace the surrounding tissue structures (phenomenon known as mass effect), which often causes herniation in the normal brain around tumor (BAT) parenchymal regions, and ultimately leads to worse prognosis \cite{chai2009field, lochhead2015etiologic, tonn2006neuro, pan2003spinal}.  While previous studies have employed radiomic (textural and shape representations) and deep-learned features obtained from within the visible peri-tumor confines alone \cite{kickingereder2016radiomic, lambin2012radiomics, shiradkar2018radiomic}, no work to our knowledge has explicitly attempted to exploit the quantifiable biomechanical deformation attributes from the BAT regions, as a complementary radiomic feature, in conjunction with features from the tumor and peri-tumoral confines. 

In this work, we presented r-DepTH, a radiomic descriptor that leverages both morphological and biomechanical attributes of the tumor regions, and employed it in the context of survival prognostication in GBM. This was achieved by combining measurements that capture subtle tissue deformation features occurring in the surrounding healthy BAT regions due to mass effect, with 3D COLLAGE descriptor, which measures local heterogeneity via higher order statistics of local gradient tensors on a voxel-wise basis, from the tumor and peri-tumoral confines. This integrated feature set was then employed to predict overall survival in GBM tumors.

Previous radiomic studies have investigated extracting features from the intra-tumor and peri-tumoral boundaries, for survival prediction or improving disease diagnosis. For instance, the study in \cite{beig2019perinodular} exploited radiomic shape and texture features extracted from both intra- and perinodular regions (where annular rings were generated around the nodule), to differentiate between cancerous lung nodules and benign masses. Similarly, another study, \cite{prasanna2017radiomic}, attempted to prognosticate survival in GBM patients using radiomic texture features extracted from the peritumoral brain parenchyma. The COLLAGE features employed within r-DepTH have previously demonstrated success in characterizing tumor heterogeneity from tumor and peritumoral regions to distinguish similar-appearing pathologies, as well as prognosticate survival, across different applications such as brain, lung, and prostate cancer \cite{shiradkar2018radiomic,prasanna2019radiomics,khorrami2019combination}. In line with our findings, most of these works have reported improvement in their diagnostic/prognostic models with inclusion of both intra and peri-lesion textural features to characterize disease heterogeneity. Our results build on these previous findings by including deformation features from brain around tumor region to the textural features from the lesion and peri-lesional compartments, to further improve the prognostic model for GBM survival.

Apart from radiomic features, a few DL approaches in literature have investigated survival prediction in GBM by interrogating features from the visible tumor confines \cite{lao2017deep,chato2017machine,nie20163d}. When replicating one such model in \cite{lao2017deep} for GBM survival prognosis on our cohort, the results did not yield significant differences between the H-R and L-R groups on the test cohort ($p$-value = $0.5$), as well as a poor C-index of $0.5$. This, we posit, could be on account of the scanner-specific variations in our multi-institutional cohort, which previously has been shown to impact the performance of CNN models across datasets from different sites and scanners \cite{yan2020mri}. 


The closest work to our work on exploring structural brain deformations was performed by \cite{prasanna2019mass}, where structural deformations were obtained from different parcellations in the brain, which were then associated with overall survival in GBM patients. Decreased survival time was found to be associated with increased deformations in certain cognitive and motor-control brain areas. Uniquely, our study found statistically significant structural deformation changes around the tumor boundaries up to $60 mm$, both in the training and the test sets, that contributed to the prognostic signature for distinguishing between high- and low-risk GBM patients. As shown in Figure \ref{fig:box},
some of our top deformation features included skewness, an indicator of lack of data symmetry, at $5 mm$ and $10 mm$.
The higher skewness values exhibited by the H-R group with poor survival (Figure \ref{fig:box}) could be linked to the way such aggressive tumors proliferate and exert pressure on BAT, and hence may lead to more lopsidedness in the frequency distribution of the deformation magnitude values at these regions, compared to the group with prolonged survival. Kurtosis, a measure of the extreme values in a dataset, at $5mm$ also turned out to be a top prognostic feature of the two risk groups, where it showed higher values for the H-R group (Figure \ref{fig:box}). This could be on account of the higher heterogeneity of BAT in patients with poor survival, due to active proliferation and herniation beyond tumor confines, leading to higher deformation magnitudes with extreme values. 

Our study did have some limitations. One limitation is that we did not explicitly account for direction (i.e phase) attributes of tissue deformations obtained in the BAT region. Further, while r-DepTH in this study was hypothesized to serve as a surrogate measure for the pressure/force exerted by the tumor, this can only be affirmed via controlled in-vivo experiments in a pre-clinical setting. Lastly, molecular and clinical information (i.e IDH, MGMT, EOR) was not available for a majority of our studies, and hence could not be used to control for clinical parameters, molecular status, and EOR, while building our prognostic risk assessment models.

\section{Conclusion}
In this work, we presented r-DepTH, an integrated radiomic descriptor that aimed at comprehensively characterizing the field effect from tumor, peri-tumor, and brain around tumor regions, towards predicting overall survival in glioblastoma patients. Our results suggest that combining measurements quantifying subtle biomechanical deformations from the brain around tumor, along with morphological features within the tumor and peri-tumor confines allowed for improved prognostic models for predicting overall survival in GBM, as compared to using clinical variables, as well as using radiomic and deep-learning features from the tumor confines alone. Future work will involve integrating the direction (i.e. phase) attributes of the tissue deformation along  with deformation magnitude features to build an integrated prognostic signature of the tumor habitat. Additionally, we  plan to extend our analysis to a large multi-site retrospective cohort, and eventually to prospectively collected scans for validation of r-DepTH as a prognostic marker for GBM and other solid tumors.




%





\ifCLASSOPTIONcaptionsoff
  \newpage
\fi

\bibliographystyle{IEEEtran}
\bibliography{IEEEabrv,Bibliography}

\vfill


\end{document}